\documentclass{mn2e}
\usepackage{epsfig}
\usepackage{longtable}
\title[Cepheid Parallaxes and the Hubble Constant]{Cepheid Parallaxes
and the Hubble Constant}
\author[van Leeuwen, Feast, Whitelock, Laney]
{Floor van Leeuwen$^{1}$, Michael W. Feast$^{2}$, 
Patricia A. Whitelock$^{2,3,4}$, Clifton D. Laney$^{3}$\\
$^{1}$ Institute of Astronomy, Madingley Rd, Cambridge, England\\
$^{2}$ Astronomy Dept., University of Cape Town, Rondebosch, 7701,
South Africa \\
$^{3}$ South African Astronomical Observatory, P.O. Box 9, Observatory,
7935, South Africa\\
$^{4}$ National Astrophysics and Space Science Programme, Department of
Mathematics and Applied Mathematics, University of
Cape \\Town, Rondebosch, 7701, South Africa\\}
\begin{document}
\maketitle
\begin{abstract}
 Revised Hipparcos parallaxes for classical Cepheids are analysed
together with 10 HST-based parallaxes (Benedict et al.). In
a reddening-free $V,I$ relation we find that the coefficient
of $\log P$ is the same within the uncertainties in our Galaxy as in
the LMC, contrary to some previous suggestions. 
Cepheids in the inner region of NGC\,4258 with near solar metallicities
(Macri et al.)
confirm this result.
We obtain a zero-point for the reddening-free relation and apply it
to the Cepheids in galaxies used by Sandage et al. to calibrate the
absolute magnitudes of SNIa and to derive the Hubble constant.
We revise their result for $H_{0}$
from 62 to $ \rm 70 \pm 5\, km\,s^{-1}\,Mpc^{-1}$.
The Freedman et al. 2001 value is revised from 72 to
$\rm 76 \pm 8\, km\,s^{-1}\,Mpc^{-1}$. 
These results are insensitive to Cepheid metallicity corrections.
The Cepheids in the inner region of NGC\,4258
yield a modulus of $29.22 \pm 0.03$ (int.) compared with an maser-based
modulus of $29.29\pm 0.15$. Distance moduli for the LMC,
uncorrected for any metallicity effects, are:
$18.52 \pm 0.03$ from a reddening-free relation in $V,I$;
$18.47 \pm 0.03$ from a period-luminosity relation at $K$;
$18.45 \pm 0.04$ from a period-luminosity-colour relation in
$J,K$. Adopting a metallicity correction in $V,I$ from Macri et al.
leads to a true LMC modulus of $18.39 \pm 0.05$.
\end{abstract}
\section{Introduction}
 The success of the Hipparcos astrometric satellite in obtaining a large
number of absolute stellar parallaxes with much greater accuracy than had
previously been possible (ESA 1997), allowed a first estimate to be made of
the zero-point of a Cepheid Period-Luminosity (PL) relation directly from
Cepheid parallaxes (Feast \& Catchpole 1997).  Investigations of the
Hipparcos data continued after the publication of the catalogue, and led to
the identification of a number of repairable problems associated with the
reconstruction of the satellite's attitude (van Leeuwen 2005). This
ultimately led to a completely new reduction of the astrometric data (van
Leeuwen \& Fantino 2005), the results of which are soon to be published (van
Leeuwen 2007) and have been incorporated in the current study. The main
impact of the reductions is for the brightest stars, where improvements of
up to a factor of four in parallax accuracy can be reached. For many of the
Cepheids improvements by up to a factor of two have been achieved.

In the present paper we discuss and analyse the revised Hipparcos parallaxes
of (classical) Cepheids. Recently, the parallaxes of 10 Cepheids, measured
with the Hubble Space Telescope (HST) and corrected to an absolute reference
frame using ground-based observations, have been published 
and discussed (Benedict et al.
2002, Benedict et al. 2007 henceforth FB2007) and we have incorporated these
data in our analysis.
 
Since the discussion of the original Hipparcos data there has been much work
on Cepheid photometry, period-luminosity (PL) relations, metallicity effects
etc. and we have been able to take advantage of this in the present work.
There has also been a great deal of theoretical work on Cepheids. We do not
discuss this since our aim has been to obtain empirically-based conclusions.
We establish a reddening-free PL relation in $V$ and $I$ (PL($W$)) as well
as PL and Period-Luminosity-Colour (PLC) relations in $J$ and $K$. These
relations should be of use in a variety of contexts both Galactic and
extragalactic and for testing theoretical models. In the present paper we
confine ourselves to the implication of our results for Cepheid-based
distances of galaxies including the LMC and the estimation of the Hubble
constant.

\section{Data}
The basic data used here are tabulated in the appendix
together with notes on individual stars. The table contains only
those stars which had all the data required for the analyses
and which are considered to be classical Cepheids.

Table A1 contains:\\
{\bf 1. Hipparcos number (no).}\\
{\bf 2. Hipparcos parallax ($\pi$).} Here and throughout in
milliarcsec (mas)\\
{\bf 3. Hipparcos parallax standard error ($\sigma_\pi$).} 
In mas.\\
{\bf 4. Star name.}\\
{\bf 5. Logarithm of fundamental period.}
Where the star is an overtone pulsator the fundamental period, $P_{0}$,
was derived from the observed period, $P_{1}$, using the relation
(Alcock et al. 1995, Feast \& Catchpole 1997):
\begin{equation}
 P_{1}/P_{0} = 0.716 -0.027\log P_{1}.
\end{equation}
Such stars are denoted by ``O" in the notes.
Here and throughout the periods are in days.\\
{\bf 6,7,8. Intensity mean magnitudes $<B>,<V>,<I>$.}
The $I$ magnitudes are on the Cousins system.\\
The $BVI$ photometry used was from Berdnikov (private communication 2005) 
and is
an update of Berdnikov et al. (2000). Where values of $I$ were not
available in this source they were taken from Groenewegen (1999)
who transformed other workers' data to the Cousins system. These
stars are marked ``G" in the notes. \\
{\bf 9,10,11. Intensity mean magnitudes $<J>,<H>,<K>$ on the SAAO system 
(Carter 1990).}
 Where possible these are from multiple observations. They are mainly from 
the papers by Laney \& Stobie (1992, 1993, 1994) and previously unpublished 
SAAO data. The individual observations on which these latter intensity
means depend will be published separately.
In a limited number of cases intensity means are from
Groenewegen's (1999) tabulation. All these stars are indicated 
by ``L" in the 
notes. Where no source is specified the mean  magnitude has 
been estimated from 
the single 2MASS measure transformed to the SAAO system (Carpenter 2001)
and corrected for phase by the procedure outlined by
Soszy\'{n}ski et al. (2005).  Since
in many cases the phases used are old (GCVS) the accuracy of these
corrections can be poor and this is taken into account in the later 
work.\\
{\bf 12. $ET$ is the value of $E(B-V)$ given by Tammann et al. (2003).}
If this is not available then this column contains 
$EF \times 0.951$ (see Tammann et al. 2003) and is indicated by ``F" in the
notes.\\
{\bf 13. $\Delta$ is the full $V$ amplitude of the star}.\\
{\bf 14. $EF$ is the value of $E(B-V)$,} as estimated by Fernie at al. (1995)
and given in the DDO data base as $FE1$. If this is not available
$FE2$ is given.\\
{\bf 15. Notes using the following symbols:}\\
$O?$, possible overtone, but considered a fundamental pulsator.\\
$B, B?$, binary or possible binary.\\
$VB$, visual binary.\\
$SB$, spectroscopic binary.\\
$SB2$, spectroscopic binary with both spectra observed.\\
The data on the binaries are mainly from the Szabados data base 
(see Szabados 2003). In cases of specific stars, additional
references are in the notes. Many of the binaries listed by Szabados are,
or possibly are, single line spectroscopic binaries. In the case of
the original Hipparcos data the effects of binaries on the
zero-point was discussed in Feast (1998)
both as regards the photometry and the astrometry. Where the photometry
may have been affected by a companion this is mentioned in the notes
and if thought appreciable the star was omitted from the analysis.\\
$DM$, double mode pulsator; the fundamental period is listed.\\
{\bf 16. The type of astrometric solution using the following codes:}\\
5: standard 5-parameter solution for single stars.\\
25: standard 5-parameter solution for a variable double star.\\
65: standard 5-parameter solution for the photo-centre of a
variable double star.\\
105: standard 5-parameter solution for the secondary in a
resolved binary with a variable component (a difficult solution).\\
7: single star with time-dependent proper motion
(accelerated solution, which may indicate long-term orbital motion).\\
3: variability induced mover, a probably spurious 
indication of duplicity
depending on the phase of the lightcurve.\\
1: stochastic solution, too much unexplained noise left in the
data, generally unreliable, and an indication of short-term
orbital motion.\\

Initial tests showed that with the Hipparcos parallaxes and listed
photometry, the following stars lay 5$\sigma$ or more from any reasonable PL
relation: Y Sgr, V350 Sgr, GQ Ori, SY Nor, HL Pup.  The Hipparcos data for
all of these stars were omitted from the analysis, although only Y Sgr would
have sufficient weight to make a significant contribution to the solutions
discussed. In some cases erroneous classification may be the cause of the
discrepancy. The reason for the significant discrepancy in the case of Y Sgr
is not fully understood since the star has an HST parallax (FB2007) in good
accord with other Cepheid data. In the combined solutions discussed below we
use only the HST parallax for Y Sgr.

Leaving aside Y Sgr there are nine Cepheids in common with the HST
parallax work. These parallaxes are listed with the Hipparcos
results in Table 1 and the weighted means are also shown. In
the solutions below we use these weighted means unless otherwise
indicated.

\begin{table}
\centering
\caption{Hipparcos and HST Cepheids in common.}
\begin{tabular}{rrrcc}
\hline
\multicolumn{1}{c}{Hipp}  & \multicolumn{1}{c}{Name} &
\multicolumn{1}{c}{$\pi_{Hipp}$} & $\pi_{HST}$ & $\pi_{mean}$\\
\hline
47854   & $l$ Car     & $2.06 \pm 0.27$ &  $2.01 \pm 0.20$ &  $2.03 \pm 0.16$\\ 
34088   & $\zeta$ Gem & $2.71 \pm 0.17$ &  $2.78 \pm 0.18$ &  $2.74 \pm 0.12$\\
26069   & $\beta$ Dor & $3.64 \pm 0.28$ &  $3.14 \pm 0.16$ &  $3.26 \pm 0.14$\\
88567   & W Sgr       & $2.59 \pm 0.75$ &  $2.28 \pm 0.20$ &  $2.30 \pm 0.19$\\
87072   & X Sgr       & $3.39 \pm 0.21$ &  $3.00 \pm 0.18$ &  $3.17 \pm 0.14$\\
110991   & $\delta$ Cep & $3.81 \pm 0.20$ & $3.66 \pm 0.15$ &  $3.71 \pm 0.12$\\
93124   & FF Aql      & $2.05 \pm 0.34$ &  $2.81 \pm 0.18$ &  $2.64 \pm 0.16$\\
102949   & T Vul       & $2.31 \pm 0.29$ &  $1.90 \pm 0.23$ &  $2.06 \pm 0.22$\\
30827   & RT Aur      & $-0.23 \pm 1.01$ &  $2.40 \pm 0.19$ & $2.31 \pm 0.19$\\
   &             &                 &                   &\\
89968   & Y Sgr       & $3.73 \pm 0.32$  &  $2.13 \pm 0.29$ & \\
\hline
\end{tabular}
\end{table}

\section{Methods using $VI$ photometry}
 The existence of a Cepheid PL relation goes back of course to Leavitt
(1908, 1912). It is known that the PL($V$) relation in the LMC has
significant width and that this is greatly reduced, probably to within the
observational errors, by the use of a period-luminosity-colour (PLC)
relation (e.g. Martin et al. 1979). However, the value of the coefficient of
the colour term and whether it varies with period has been a matter of
uncertainty and debate.

In recent times there has been considerable discussion on the possibility
that some PL relations might be non-linear. Thus, it has been suggested
that the PL($V$) relation in the LMC is non-linear
(Sandage et al. 2004, Ngeow et al. 2005). 
However, it is not entirely clear whether 
this effect is real or, for instance,
due to systematic errors in the adopted reddenings
varying with period. In any case, there
seems to be good evidence (Ngeow \& Kanbur 2005) that in the LMC, a
 ``reddening-free"
relation (Madore 1976) such as,
\begin{equation}
M_{W} = \alpha \log P + \beta (V - I) + \gamma,
\end{equation}
is linear. Here the coefficient, $\beta$, is an adopted
ratio of total to selective extinction ($A_{V}/(A_{V} - A_{I})$). 
We have restricted our discussion at optical
wavelengths to a relation of this form because of this and
because of its importance in extragalactic work.

In a very detailed analysis of the data available to them, Sandage et al.
(2004) have suggested that the slopes of PL relations in the optical, and
hence the $\log P$ coefficient in equation 2, differ between the LMC and our
Galaxy, presumably due to metallicity effects. In the case of the Galaxy the
slopes were derived by combining Cepheid distances derived from
Baade-Wesselink (pulsation parallax) type analyses with those from Cepheids
in clusters with distances from main-sequence fitting. Their results have
remained controversial for the following reasons. Gieren et al. (2005)
showed that the Baade-Wesselink type distances of LMC Cepheids were a
function of period if a conventional $``p"$ factor was used to convert
observed radial velocities to pulsational velocities of the stars. Changing
$``p"$ to remove this effect brings the LMC and Galaxy PL slopes into
agreement within the uncertainties. As regards the `cluster' distances,
these partly depend, especially at longer periods, not on clusters but on
stellar associations. The distances of these associations remain somewhat
uncertain and their inclusion can affect any derived PL slope (as can be
inferred from the early work of Feast \& Walker 1987, their table 3). In
view of these various uncertainties, one of our aims has been to derive the
PL($W$) slope in our Galaxy.

Two PL($W$) relations are of particular importance for extragalactic
applications.\\
(1) The relation adopted by Freedman et al. (2001)  in their
HST key project on the Cepheid calibration of the Hubble Constant
($H_{0}$), 
\begin{equation}
 M_{W} = -3.255 \log P + 2.45 (V - I) - 2.724 .
\end{equation}
The $\log P$ coefficient was derived from OGLE LMC data (Udalski et al.
1999) and the colour coefficient from their adopted reddening law. The
equation as given is applicable to  local Galactic Cepheids.
The zero-point is
derived from their adopted LMC modulus and metallicity correction.\\ 
(2) The relations for Galactic Cepheids 
used by Sandage et al. (2006 henceforth
S2006)\footnote{This is the final paper of a series} in their work on a
Cepheid-based $H_{0}$ are equivalent to:
\begin{equation}
M_{W} = -3.746 \log P + 2.563(V - I) - 2.213 .
\end{equation}
The basis on which this equation was derived was
discussed above.\\

A third 
reddening-free relation is of relevance. This was derived (FB2007) from a
``cleaned" selection of 581 LMC OGLE Cepheids and adopting the
Freedman et al. reddening law
\begin{eqnarray}
M_{W} = -3.29 (\pm 0.01)\log P + 2.45 (V - I) - \nonumber\\
 (Mod -15.94 (\pm 0.01)) ,
\end{eqnarray}
where $Mod$ is the distance modulus of the LMC and no metallicity
correction has been applied.

After making corrections for reddening, Udalski et al. (1999) 
derived a true PLC relation for LMC Cepheids in the OGLE data base, 
which can be written as,
\begin{eqnarray}
V_{0} = -3.25(\pm 0.02) \log P + 2.41 (\pm 0.03) (V-I) + \nonumber\\
 15.88(\pm 0.02)).
\end{eqnarray}
Comparison of equations 5 and 6 shows that the coefficients of
the slopes and colour terms are nearly the same in the two equations.
Thus, we are justified (at least in the LMC) in considering
a reddening-free relation as also very close to a true PLC
relation. This is important for two reasons. First, as discussed above,
we expect a PLC relation to be very narrow. Secondly, because Cepheid
overtone pulsators are in general bluer than those of the same
fundamental period, they lie systematically above fundamental
pulsators in a PL($V$) plot. However, they lie together
with the fundamental pulsators in a PLC plot (see, e.g.
Beaulieu et al. 1995).
This is important if we wish to include overtone pulsators in
a Cepheid calibration. We test this result in the case of
Galactic Cepheids below.

In analysing the data we have proceeded in two ways:\\
(1) For a limited number of Cepheids the percentage errors in
the parallaxes are sufficiently small that individual values of
$M_{W}$  can be directly combined to derive PL relations; Cepheids
selected in this way require a Lutz-Kelker type bias correction
(Lutz \& Kelker 1973).
We have scaled these corrections to be on the same system as that
adopted by FB2007. We are primarily interested in using this subset of
stars to obtain an estimate of $\alpha$ in equation 2.\\
(2) We combine the main body of data using the method of reduced parallaxes
(e.g. Feast 2002) 
and fixed values of $\alpha$ to obtain the zero-point,
$\gamma$, in equation 2.

\section{The slope $\alpha$ of the Galactic PL($W$) relation}
  Table 2 lists the subset of Cepheids used in this section. The table
gives, Hipparcos number, name, 
the parallax and its standard error from the combined Hipparcos and HST
data, 
the absolute magnitude, $M_{W}$, (without
Lutz-Kelker correction) and its standard error (= $2.17\sigma_{\pi}/\pi$), 
from this parallax and the adopted photometry together
with $\beta = 2.45$. Also given are the Lutz-Kelker corrections applicable
in this case and the
$\log$ of the fundamental period. Fig.~1 shows $M_{W}$
with Lutz-Kelker corrections applied plotted against $\log P$
and with our finally adopted relation ($\alpha =-3.29$,
$\beta =2.45$ and $\gamma = -2.58$) shown. The residuals in this case
are also listed in Table 2.
\begin{figure}
\centerline{\psfig{figure=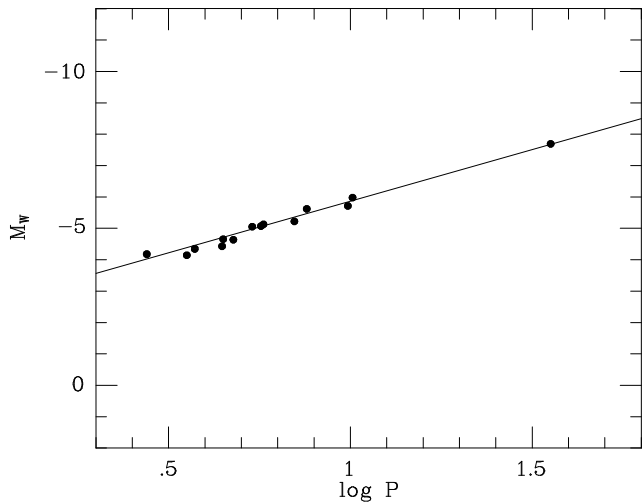,width=84mm}}
\caption{$M_{W}$ (with Lutz-Kelker correction) plotted against $\log P$ 
for the 14 stars in Table 2. The line is the relation finally adopted 
which has $\alpha  =-3.29$, $\beta = 2.45$ and $\gamma = -2.58$}
\end{figure}

\begin{table*}
\centering
\caption{Cepheids used in the determination of the PL(W) slope
$\alpha$.}
\begin{tabular}{rrrrrrr}
\hline
\multicolumn{1}{c}{Hipp} & \multicolumn{1}{c}{Star} & 
\multicolumn{1}{c}{$\pi $} & $\log P$ 
& \multicolumn{1}{c}{$M_{W}$} & \multicolumn{1}{c}{LK Corr.} &Res.\\
\hline
11767   & $\alpha$ UMi & $7.72 \pm 0.12$ &  0.754 & $ -5.08 \pm 0.03$ 
&   0.00 & --0.02\\
13367   & SU Cas & $2.57 \pm 0.33$     & 0.440 & $ -4.18 \pm 0.28$ 
& --0.13 & --0.15\\
26069   & $\beta$ Dor  & $3.26 \pm 0.14$    & 0.993 & $ -5.72 \pm 0.09$ 
& --0.02 & +0.13\\
30827   & RT Aur       & $2.31 \pm 0.19$   & 0.572 & $ -4.35 \pm 0.18$ 
& --0.06 & +0.11\\
34088   & $\zeta$ Gem  & $2.74 \pm 0.12$    & 1.006 & $ -5.98 \pm 0.10$ 
& --0.02 & --0.09\\
47854   & $l$ Car      & $2.03 \pm 0.16$      & 1.551 & $ -7.70 \pm 0.17$ 
& --0.05 & --0.01\\
61136   & BG Cru       & $2.23 \pm 0.30$    & 0.678 & $ -4.63 \pm 0.29$ 
& --0.15 & +0.18\\
87072   & X Sgr        & $3.17 \pm 0.14$   & 0.846 & $ -5.22 \pm 0.10$ 
& --0.02 & +0.14\\
88567   & W Sgr        & $2.30 \pm 0.19$    & 0.880 & $ -5.62 \pm 0.18$ 
& --0.06 & --0.15\\
89968   & Y Sgr        & $2.13 \pm 0.29$  & 0.761 & $ -5.13 \pm 0.30$ 
& --0.15 & --0.05\\
93124   & FF Aql       & $2.64 \pm 0.16$    & 0.650 & $ -4.66 \pm 0.13$ 
& --0.03 & +0.06\\
102949   & T Vul        & $2.06 \pm 0.22$     & 0.647 & $ -4.43 \pm 0.23$ 
& --0.09 & +0.28\\
104185   & DT Cyg       & $2.19 \pm 0.33$  & 0.550 & $ -4.15 \pm 0.33$ 
& --0.18 & +0.24\\
110991   & $\delta$ Cep & $3.71 \pm 0.12$    & 0.730 & $ -5.05 \pm 0.07$ 
& --0.01 & --0.07\\
\hline
\end{tabular}
\end{table*}
Weighted least 
square fits to equation 2 were made to various subsets of the data
and the slopes ($\alpha$) derived are listed in Table 3.
This is in two parts: Table 3(a) adopts $\beta$ = 2.45 (as in
Freedman et al. 2001 ) and Table 3(b) adopts $\beta$ = 2.523 (as in 
S2006) (see section 3). 
\begin{table}
\centering
\caption{Determinations of the slope, $\alpha$, in equation 2.}
\begin{tabular}{llll}
\hline
No & Sample & $\alpha$ &  $\beta$\\
\hline
 & & (a) &\\
\hline
1 & 10 HST stars + HST phot. & $-3.335 \pm 0.172$ & 2.45\\
2 & 10 HST stars + New phot. & $-3.473 \pm 0.183$ & 2.45\\
3 & 10 stars HST + Hipp      & $-3.328 \pm 0.188$ & 2.45\\
4 & (3) + Polaris (11 stars) & $-3.285 \pm 0.169$ & 2.45\\
5 & weight $>10$ (13 stars)  & $-3.273 \pm 0.155$ & 2.45\\
6 & weight $>8$  (14 stars)  & $-3.288 \pm 0.151$ & 2.45\\
7 & (6) omitting $l$ Car (13 stars) & $-3.265 \pm 0.230 $ & 2.45\\
8 & LMC (OGLE)               & $-3.29 \pm 0.01 $ &  2.45\\
9 & Freedman (adopted)       & $-3.26          $ &  2.45\\
\hline
& & (b) &\\
\hline
10 & 10 HST stars + New phot. & $-3.502 \pm 0.182$ & 2.523\\
11 & 10 stars HST + Hipp      & $-3.357 \pm 0.186$ & 2.523\\ 
12 & (11) + Polaris (11 stars)& $-3.330 \pm 0.165$ & 2.523\\
13 & weight $>$ 10 (13 stars) & $-3.315 \pm 0.152$ & 2.523\\
14 & weight $>$ 8 (14 stars   & $-3.330 \pm 0.149$ & 2.523\\
15 & Sandage (adopted)        & $-3.75           $ & 2.523\\
\hline
\end{tabular}
\end{table}

From Table  3(a) we draw the following conclusions:\\
1. Slight changes in the adopted photometry affect the value of the slope.
However, this is not the main source of uncertainty.\\
2. Adding the overtone Polaris at its fundamental period does not
change the slope appreciably (see also the next section).\\
3. Leaving out $l$ Car does not affect the slope appreciably. This 
is important since $l$ Car is by far the longest period star in
this sample and therefore, when included, has a major effect on
the slope determined.\\
4. Our best determinations (solutions 6 and 7 of Table 3) give values of
$\alpha$ close to that determined for LMC stars from the OGLE
data (as shown in the table). There is still significant
uncertainty in our value of the Galactic slope. However, within
the uncertainties it agrees with that determined for the LMC.\\
From Table 3(b) we draw the following conclusions:\\ 
1. Using the ``Sandage" colour
coefficient ($\beta$) we get slopes which are not significantly different
from those in Table 3(a).\\
2. Our slopes, especially the higher weight ones, are distinctly
different from that adopted by S2006, and in view of the
uncertainties surrounding this latter result (see section 3) we 
suggest it should be replaced by our best value. Nevertheless,
in the next section we give zero-points derived using  
a value of $\alpha = -3.75$.\\

The main conclusion of this section is that within current uncertainties the
value of $\alpha$ is the same in the Galaxy as in the LMC and
our final results will be based on this assumption.

\section{The zero-point $\gamma$ of the PL($W$) relation}
\begin{table}
\centering
\caption{The zero-point, $\gamma$, of equation 2  with fixed $\alpha$ and
$\beta$.}
\begin{tabular}{llllll}
\hline
No.& Stars & $\alpha$ & $\beta$ & $\gamma$ & notes\\
\hline
& & & (a) & &\\
\hline
1 & 14 & --3.255 & 2.45 & $-2.606 \pm 0.022$ & \\
2 & 10 & --3.255 & 2.45 & $-2.568 \pm 0.036$ & HST result\\ 
3 & 14 & --3.29  & 2.24 & $-2.579 \pm 0.022$ & \\
4.& 14 & --2.75  & 2.523 & $-2.264 \pm 0.028$ &\\
\hline
& & & (b) & &\\
\hline
5 & 240 & --3.255 & 2.45 & $-2.604 \pm 0.030$ &\\
6 & 240 & --3.29  & 2.45 & $-2.576 \pm 0.030$ &\\
7 & 239 & --3.29  & 2.45 & $-2.558 \pm 0.044$ & (6) no Polaris\\
8 & 213 & --3.29  & 2.45 & $-2.563 \pm 0.046$ & (6) no overtones\\
9 & 240 & --3.75  & 2.523 & $-2.263 \pm 0.030$ &\\
\hline
\end{tabular}
\end{table}

Table 4 gives results of the determinations of the zero-point,
$\gamma$ in equation 2 using fixed values of $\alpha$ and
$\beta$. In Table 4(a) are the values of $\gamma$ obtained
directly from the 14 stars in Table 2 with Lutz-Kelker
corrections applied. In Table 4(b) we give the values
of $\gamma$ derived by the method of reduced parallaxes 
(e.g. Feast 2002)
to our bulk sample.
Points to note are:\\
1. Values of $\gamma$ in Table 4(a) agree closely with the
corresponding values in Table  4(b).\\
2. In Table 4(b), leaving out the high weight overtone 
pulsator, Polaris, makes no significant difference. Nor does
leaving out all the known overtone pulsators. This is 
consistent with the discussion in section 3 which noted
that the reddening-free relation in the LMC was very closely
the same as a true PLC relation.\\
3. In carrying out the reduced parallax solutions we have assumed that
 the uncertainties are dominated by the errors in the parallaxes.
That is, we have neglected the second term in equation 3 of Feast (2002). 
However, if we supposed that $\sigma_{m_{0}}$ = $\sigma_{M_{0}}$ = 0.14,
which we believe would be a gross overestimate, then 
the value of $\gamma$ in Table 4(b) (solution 6) would only change from
--2.576 to --2.554.

We adopt --2.58 (solution 6, Table 4)
as the value for $\gamma$ to use with $\alpha = -3.29$
and $\beta = 2.45$ in equation 2. 

\section{The Hubble Constant}
An important use of Cepheids is as a basis for the determination
of the distances of galaxies and from that 
the estimation of a value of $H_{0}$. This 
can then be compared with the value derived in other ways
(e.g. from the microwave background) which require the adoption
of a general cosmological model; thus providing a test of the model.
 Sandage and his co-workers have recently completed a major programme of
reanalysing HST data of Cepheids in galaxies in which supernovae have been
observed (see S2006, Saha et al. 2006 and earlier papers in the series). 
In their summary paper they
use the Cepheid data to derive distance moduli to 10 normal SNIa.
From these they derive the maximum SNIa brightness (as defined
by them). They then use this as a zero-point for a determination
of $H_{0}$. We have redetermined the distance moduli of
these galaxies using equation 2 with 
our adopted coefficients from section 5
viz. $\alpha = -3.29$, $\beta = +2.45$,
and $\gamma = -2.58$. We use the same selection
of Cepheids as used by S2006 and adopt the corrected apparent
SN magnitudes in Table 2 of that paper. We derive SNIa absolute
magnitudes equivalent to those in Table 3  of S2006
and like them derive weighted means. Table 5
gives the weighted mean absolute magnitudes of S2006 and the 
corresponding values of $H_{0}$ which this implies in their work.
\begin{table*}
\centering
\caption{SNIa absolute magnitudes and $H_{0}$.}
\begin{tabular}{lllll}
\hline
 & $M_{B}$ & $M_{V}$ & $M_{I}$ &\\
& $H_{0}$ (B) & $H_{0}$ (V) & $H_{0}$ (I)& Adopted $H_{0}$\\
\hline
 S2006 & $-19.49 \pm 0.04$ & $-19.46 \pm 0.04$ & $-19.22 \pm 0.05$ &\\
&$62.4 \pm 1.2$ & $62.5 \pm 1.2$  & $62.1 \pm 1.4$ & $62.3 \pm 1.3$ (int.)\\
\hline
Revised (no metal cor.)& $-19.26 \pm 0.05$ & $-19.22 \pm 0.05$ &
$-18.98 \pm 0.07$ &\\
& $69.4 \pm 1.6$ & $69.8 \pm 1.6$ & $69.4 \pm 2.3$ & $69.5$\\
\hline
Revised (with metal cor.) & $-19.26$ & $-19.22$ & $-18.97$ &\\ 
& $69.4$ & $69.8$ & $69.7$ & $69.6$\\
\hline
\end{tabular}
\end{table*}

Table 5 also contains the equivalent weighted mean absolute 
magnitudes derived
from our estimates of the moduli. Evidently the difference between
our absolute magnitude and theirs implies a change in $H_{0}$
and this is given in the table
(in units of $\rm km\,s^{-1} \,Mpc^{-1}$). The table contains the results
of two estimates we have made for the moduli. In one case we have applied
no metallicity correction. In the other we have applied a metallicity
correction based on the ``Sakai" values of [O/H] in table 1 of 
S2006. These abundances are in the $T_{e}$ system and
Macri et al. (2006) find from their work on NGC\,4258 that in this system
a PL($W$) relation such as we have used requires a correction of
--0.49 mag/dex. There is considerable uncertainties in the size of
the required metallicity correction. Fortunately, as Table 5
shows, the result we obtain is quite insensitive to the correction
used. This is due to the mean metallicity of the
S2006 galaxies being close to that of the local Cepheids.
Only a large, non-linear metallicity correction would
change this conclusion.
 
Thus our best value of $H_{0}$ based on our PL($W$) relation but with 
all other data and assumptions as in S2006 is 69.6.
S2006 obtained $62.3 \pm 5$. Our improved Cepheid results would
in principle reduce the uncertainty
to $\sim \pm 2$, but to be conservative we keep it
the same. Our revised value ($70\pm 5$) is clearly compatible with the value
of $73 \pm 3$ found from the WMAP data and a $\Lambda$CDM model
(Spergel et al. 2006).

There are a large number of other determinations of $H_{0}$, some of them
depending on a Cepheid scale. The most widely quoted is that of Freedman et
al. (2001) who obtained $H_{0} = 72 \pm 8$.  Since that paper was published
there has been much work on the refinement of the basic HST data, on galaxy
metallicities and on the various large scale distance indicators used by
these workers. However, the fact that the values of $\alpha$ and $\beta$ in
the PL($W$) relation they use are close to ours, and that the mean
metallicity of their sample, weighted according to the contribution of an
indicator to $H_{0}$ is close to solar, means that a satisfactory estimate
of the effect of our work on theirs can be made by comparing PL($W$)
zero-points. Table 4 solution 5 shows that with their $\alpha$ and 
$\beta$ we find
$\gamma = -2.604$ whereas they used $-2.724$ 
at the metallicity
of Galactic Cepheids (see equation 3 above).
The difference (0.12 mag) implies an increase of their $H_{0}$ to $76 \pm
8$, where to be conservative we keep the error the same,
though the discussion of Freedman et al. together with our new results
would in principle reduce this to $\sim \pm 6$.
Again this revised $H_{0}$ is
quite compatible with the WMAP result.

\section{The distance modulus of the LMC using $VI$ photometry}
   Combining the LMC PL($W$) relation (equation 5) with our derived value
of $\gamma$ ($-2.58$) gives directly the true modulus of the LMC
uncorrected for metallicity effects. We thus find a modulus of $18.52\pm
0.03$. Adopting the results of Andrievsky et al. (2002) and Sakai et al.
(2004), as discussed by S2006 the LMC Cepheids are metal deficient by $\Delta
[O/H] = 0.26$ on the ``$T_{e}"$ abundance scale. As already noted Macri et
al. (2006) found a Cepheid metallicity effect, applicable to our PL($W$)
results, of $-0.49 (\pm 0.15)$ mag/dex. Applying this leads to a metallicity
corrected LMC modulus of $18.39 \pm 0.05$.

The main uncertainty in this result is due to the uncertainty in the
metallicity correction to our PL($W$) relation (note that corrections at
other wavelengths would not necessarily be the same). It has even been
recently suggested that the effect may be negligible (Rizzi et al. 2007).
This would be somewhat remarkable since it has been long known (Gascoigne \&
Kron 1965 and much further work) that the intrinsic colours of LMC Cepheids
differed from those of Galactic Cepheids of the same period. This was shown
by Laney \& Stobie (1986) to be due to a combination of changes in
atmospheric blanketing together with a real shift of the instability strip
in temperature. Fortunately the metallicity correction problem is not
important for the work on $H_{0}$ discussed in the previous section. The use
of the LMC modulus, however determined, will remain an uncertain basis for
an extragalactic scale based on Cepheids pending further work on the
metallicity effect.

\section{The Cepheid-based distance of the Maser-host galaxy NGC\,4258}
NGC\,4258 is of special interest because a distance has been determined
(Herrnstein et al. 1999) based on the motions of $H_{2}O$ masers around the
central black hole. Macri et al. (2006) have obtained HST photometry
of Cepheids in this galaxy. Here we concentrate on Cepheids in their inner
field which has a metallicity close to solar and therefore is
immune to the problem of metallicity corrections when combined with our
Galactic calibration. Some discussion of the Cepheid-based distance of this
galaxy was made in connection with the HST Cepheid parallax work (FB2007).

We first consider the coefficient $\alpha$ of equation 2.  Taking the 69
Cepheids which pass the selection criteria of Macri et al.\footnote{ Of the
74 Cepheids which pass their stated selection criteria, the following stars
were rejected by them as outliers: I-040434, kI-008361, I-144755, I-081614,
I-I-009241 (L. Macri private communication).}
 We find the following:\\
1. For our adopted value of $\beta = 2.45$ we find
 $\alpha = -3.18 \pm 0.13$.\\
2. For the value of $\beta = 2.523$ favoured by S2006
we find $\alpha = -3.19 \pm 0.13$.\\
These values are less than $1 \sigma$ from our adopted slope of
--3.29, and $4.3 \sigma$ different from that favoured by S2006
for Galactic Cepheids (--3.75). We take this as further evidence
that the slope of the PL($W$) relation in the LMC applies also to
Cepheids of approximately Galactic composition. Macri et al. reach a
similar conclusion by a different method. 

Adopting $\alpha = -3.29$, $\beta = 2.45$ and $\gamma = 2.58$, from section
5,
for equation 2 we find from the data of Macri et al. (2006) for their inner
region, a distance modulus of $29.22 \pm 0.03$ for NGC\,4258. The standard
error takes into account the internal scatter in the NGC\,4258 Cepheid data
and the uncertainty in the adopted $\gamma$. The currently available
maser-based distance modulus is $29.29 \pm 0.15$ (Herrnstein et al. 1999) 
and
is thus in good agreement with the parallax-based Cepheid modulus.
The referee has suggested that the Cepheid modulus may be slightly
underestimated due to the possible effects of blending on the
NGC4258 Cepheids and an even closer agreement with the maser distance
might be obtained if this could be taken into account.
Improvements in the maser-based modulus are expected (Macri et al. 2006,
Argon et al. 2007) and these should allow a more stringent comparison with
the Cepheids.

\section{Methods using $JK$ photometry}
 In the present section our aim is to establish zero points for PL and PLC
relations in the near infrared. The data used were outlined in section 2 and
listed in Table A1.  In the case of the important overtone pulsator,
Polaris, the 2MASS observation is heavily saturated and thus has a very
large uncertainty and there appears to be no other ground-based near
infrared photometry on a system which can be reliably transformed to the
SAAO system. There is, however, extensive DIRBE data (Smith et al. 2004).
This has been transformed to the SAAO system as follows. There are eleven
SAAO $JHKL$ standard stars in the DIRBE catalogue with $J <1.51$ and $K<
1.00$ and no indication of confusion in the DIRBE flags.  Using these as
calibrators, together with Procyon which has $J=-0.40$ and $-0.65$ on the
SAAO system (transformed from Glass (1974) and Bouchet et al. (1991)) leads
to $J=0.98$ and $K=0.60$ for Polaris. 

In view of the
$VI$ results of the present paper and those at $VI$ and $K$ in FB2007,
we confine ourselves to determining the zero-points of PL and PLC
relations of forms established in the LMC. The most extensive
data set in $J$ and $K$ for LMC Cepheids, 
based on multiple observations, is that of Persson et al. (2004).
We adopt their PL and PLC relations. Transformed (Carter 1990) onto
the SAAO system these are:
\begin{equation}
 M_{K} = -3.258 \log P + \gamma_{1}
\end{equation}
and
\begin{equation}
M_{K} = -3.457 \log P + 1.894(J-K)_{0} + \gamma_{2} .
\end{equation}
We have carried through the calculations with reddening corrections
according to both the reddening law derived by Laney \& Stobie (1993) and
that of Cardelli, Clayton \& Mathis (1989) and taking values of $E(B-V)$
from either Fernie et al. (1995) or Tammann et al. (2003) as outlined in
section 2. The differences in the results obtained were all very small and
much smaller than the standard errors of the quantities sought, so we list
only one of them (Tammann et al. reddenings, Laney \& Stobie reddening law).

In the case of a PL relation we expect the overtones, at their fundamental
periods, to be brighter than fundamental pulsators of the same period,
because of a temperature difference, and this is clear from the PL($K$)
relations in the LMC by, e.g. Groenewegen (2000).  We must therefore treat
fundamental pulsators and overtones (at their fundamental periods)
separately in discussing equation 7. Note that because of the relation
between fundamental and first overtone periods (equation 1) the PL relation
when transformed from the fundamental to the observed, overtone, periods
will have a somewhat different slope, as found by Groenewegen.  Using the
method of reduced parallaxes as outlined earlier we obtain the results
listed in Table 6 where Polaris is treated separately. Using the whole
sample or only those with well covered light curves makes no significant
difference to the results.
\begin{table}
\centering
\caption{PL($K$) zero-point, $\gamma_{1}$, determinations.}
\begin{tabular}{ccc}
\hline
sample & stars & $\gamma_{1}$\\
\hline
Fundamentals & 220 & $-2.40\pm 0.05$\\
Overtones (not Polaris) & 26 & $-2.33 \pm 0.15$\\
Polaris &  1 & $-2.50 \pm 0.03$\\
\hline
\end{tabular}
\end{table}

The difference in $\gamma_{1}$ between Polaris (at its fundamental period)
and the mean of the fundamental pulsators ($-0.07 \pm 0.06$ mag) is not
significant. However, it in fact agrees closely with the difference between
overtones (at their fundamental periods) and fundamental pulsators expected
(--0.08 mag) at the period of Polaris from the LMC data of Groenewegen
(2000).

Results for the PLC zero-point are given in Table 7.

\begin{table}
\centering
\caption{Determinations of the PLC($J,K$) zero-point, $\gamma_{2}$.}
\begin{tabular}{ccc}
\hline
sample & stars & $\gamma_{2}$\\
\hline
All except Polaris & 246 & $-3.01 \pm 0.05$\\
Fundamentals & 220 & $-3.02 \pm 0.05$\\
Overtones (not Polaris) & 26 & $-2.85 \pm 0.15$\\
Polaris & 1 & $-3.07 \pm 0.03$\\
\hline
\end{tabular}
\end{table}
We expect for the PLC relation that overtones 
at their fundamental periods should follow the same
relation as fundamental pulsators. The table shows that
Polaris is $+0.05 \pm 0.06$ mag fainter than
the mean of the fundamental pulsators, not significantly
different from zero.

\section{The LMC modulus from $J$ and $K$}
The LMC relations of Persson et al. (2004) converted to the SAAO system
using the relations of Carter (1990) are:
\begin{equation}
K_{0} = -3.258 \log P + 16.048
\end{equation} 
and
\begin{equation}
K_{0} = -3.457 \log P  + 1.894(J-K)_{0} + 15.402.
\end{equation}
Also Groenewegen (2000) obtained for LMC overtone pulsators in the
2MASS system\footnote{Groenewegen's relation for LMC fundamental pulsators is
in good agreement with that of Persson et al. (2004). This latter is
in the LCO system which is very close to that of 2MASS. The
PL slopes given by Ita et al. (2004) (LCO system) do not agree
well with Groenewegen or Persson et al. This may be connected with
the inclusion of LMC Cepheids with $\log P < 0.4$ in the Ita et al.
sample (Y. Ita private communication)}:
\begin{equation}
K_{0} = -3.381 \log P_{1} + 15.533.
\end{equation}
The zero-point,
$\gamma_{1}$, for fundamental pulsators in Table 6 together with equation 9
gives an LMC modulus of $18.45\pm 0.05$. Polaris, at its 
observed (overtone) period with $K=0.58$ (the value from section 9
converted to the 2MASS system using
Carpenter (2001)) and equation 11, gives $18.49\pm 0.04$ (taking into
account the uncertainty in Groenewegen's result. A straight mean of these
two values ($18.47 \pm 0.03$) is our best estimated of the PL($K$)
modulus of the LMC uncorrected for abundance effects. This is in
agreement with $18.45\pm 0.04$ derived by FB2007.

A weighted mean of the last three entries in Table 7 leads to
$\gamma_{2} = -3.05\pm 0.02$. Together with equation 10 and an estimate
of its uncertainty leads to a infrared PLC modulus of
$18.45\pm 0.04$ again without any metallicity correction.

These values may be compared with those found from $VI$ in section 7
which were $18.52\pm 0.03$ uncorrected for metallicity effects
and $18.39\pm 0.05$ with a metallicity correction from Macri et al. (2006).
These results suggest that any metallicity corrections to
the infrared relations will be small.

\section{Conclusions}
Our main conclusions based on the combined revised Hipparcos and
HST data are the following;\\
1. The coefficient of the $\log P$ term in a reddening free $V,I$ relation
is found to be the same, within the uncertainties, in our Galaxy
and in the LMC.\\
2. This result is supported by an analysis of the data of Macri et al.
(2006) for Cepheids in the inner region of NGC\,4258.\\
3. Our reddening-free $V,I$ relation applied to the Cepheids
in the inner region of NGC\,4258 leads to a modulus of $29.22 \pm 0.03$.
in agreement (but of higher accuracy than) the maser-based distance
($29.29 \pm 0.15$)\\
4. Our revised Cepheid $V,I$ calibration leads to a revision of the
Cepheid-based distances to the 10 galaxies on which S2006 base their
SNIa calibration and Hubble constant. We revise their results
from $H_{0} = 62$ to $\rm 70 \pm 5\, km\,s^{-1}\,Mpc^{-1}$
This result is immune to metallicity effects on the Cepheid scale
as long as these are linear.
The Freeman et al. (2001) result is revised from 72 to
$H_{0} = 76 \pm 8\,\rm {km\,s^{-1}\,Mpc^{-1}}$.
Both these results are consistent with the 
recent WMAP value 
($H_{0} = 73 \pm 3 \, \rm {km\,s^{-1}\,Mpc^{-1}}$)\\
5. The zero-points of Galactic PL($K$) and PLC($J,K$) are derived.\\
6. Applying our various relations to the LMC we find the
following distance moduli, uncorrected for metallicity effects:
$18.52 \pm 0.03$ from a reddening-free $V,I$ relation;
$18.47 \pm 0.03$ from a PL($K$) relation; $18.45 \pm 0.04$
from a PLC($J,K$) relation.\\
7. Applying a metallicity correction derived by Macri et al. (2006)
to our LMC modulus leads to a true (metallicity corrected)
modulus of $18.39 \pm 0.05$

\section*{Acknowledgments}
We are very grateful to Dr L. Berdnikov for placing the unpublished
revision of his catalogue of Cepheid photometry at our disposal.
Dr L. Macri and Dr Y. Ita, very kindly and promptly, answered a 
number of queries about their work. We thank the referee
(Prof. W. Gieren) for his comments

\end{document}